\documentclass[12pt]{article}
\usepackage{i_style}
\usepackage{subfigure}
\usepackage[nofiglist,notablist]{endfloat}
\usepackage{amsmath}
\usepackage{aasjournals}
\usepackage{natbib}
\usepackage{xspace}
\usepackage{graphicx}

\newcommand{\mkomma}{\mbox{,}\;}                 
\newcommand{\mpkt}{\; \mbox{.}}                   
\newcommand{\D}{\displaystyle}

\newcommand{\bmu}{\mbox{\boldmath$\mu$}}
\newcommand{\bnabla}{\mbox{\boldmath$\nabla$}}
\newcommand{\bomega}{\mbox{\boldmath$\omega$}}
\newcommand{\bB}{\mbox{\boldmath$B$}}
\newcommand{\bBT}{\mbox{\footnotesize\boldmath$B$}}
\newcommand{\bF}{\mbox{\boldmath$F$}}
\newcommand{\bM}{\mbox{\boldmath$M$}}

\newcommand{\br}{\mbox{\boldmath$r$}}

\newcommand{\bv}{\mbox{\boldmath$v$}}

\def\agg{\mathrm{agg}}
\def\eff{\mathrm{eff}}
\def\chain{\mathrm{chain}}
\def\vec#1{{\mathbf{#1}}}
\def\vec#1{{\mbox{\boldmath$#1$}}}
\def\mat#1{{\cal #1}}
\def\rotix{\ensuremath{\varrho_{i,1}}}
\def\rotiy{\ensuremath{\varrho_{i,2}}}
\def\rotiz{\ensuremath{\varrho_{i,3}}}
\def\rotit{\ensuremath{\varrho_{i,4}}}
\def\rotixdot{\ensuremath{\dot{\varrho}_{i,1}}}
\def\rotiydot{\ensuremath{\dot{\varrho}_{i,2}}}
\def\rotizdot{\ensuremath{\dot{\varrho}_{i,3}}}
\def\rotitdot{\ensuremath{\dot{\varrho}_{i,4}}}
\def\ox{\ensuremath{\omega_{i,1}}}
\def\oy{\ensuremath{\omega_{i,2}}}
\def\oz{\ensuremath{\omega_{i,3}}}
\def\RCPone{\ensuremath{\vec{P}_1}}
\def\RCPtwo{\ensuremath{\vec{P}_2}}
\def\roti{\ensuremath{\vec{\varrho}_i}}
\def\rotmat{{\cal R}}
\def\sand{{\em SAND\/}\xspace}

\begin{document}

\title{\textbf{Magnetic Aggregation I: Aggregation Dynamics and
    Numerical Modelling}}

\makeatletter

\author{Carsten Dominik\\
Sterrenkundig Instituut ``Anton Pannekoek''\\
Universiteit van Amsterdam\\
Kruislaan 403, NL-1089 SJ Amsterdam/The Netherlands\\
Email: dominik@science.uva.nl \and
Henrik N\"ubold\\
Institut f\"ur Geophysik und Meteorologie\\
Technische Universit\"at Braunschweig\\
Mendelssohnstr. 3, D-38106 Braunschweig/Germany\\
Email: h.nuebold@tu-bs.de}

\maketitle

\vfill

\noindent   number of manuscript pages: \textbf{40}\\
   including \textbf{0} tables and \textbf{5} figures\\

  \newpage
  \vspace*{3cm}
  \noindent
  RUNNING HEAD: \textbf{~~Magnetic Aggregation}\\[2cm]
  Please send editorial correspondence and proofs to:\\[1cm]
  \hspace*{1cm}\parbox{10cm}{
  Carsten Dominik\\
  Sterrenkundig Instituut ``Anton Pannekoek''\\
  Universiteit van Amsterdam\\
  Kruislaan 403\\
  NL-1089 SJ Amsterdam\\
  The Netherlands\\
  Tel. +31 20 5257477\\
  Fax  +31 20 5257484\\
  Email dominik@science.uva.nl}
  \makeatother 
  \newpage


\newpage

\begin{abstract}
    Focussing on preplanetary grains growth, we discuss the
    properties of dust aggregation driven by magnetic dipole forces.
    While there is no direct evidence for the existence of magnetic
    grains present in the solar nebula, there are reasons to assume
    they may have been present.  We derive analytical expressions for
    the cross-section of two interacting dipoles.  The effective cross
    section depends upon the strength of the magnetic dipoles and the
    initial velocities.  For typical conditions the magnetic cross
    section is between 2 and 3 orders of magnitude larger than the
    geometric cross section.  We study the growth dynamics of magnetic
    grains and find that the mass of the aggregates should increase
    with time like t$^{3.2}$ whereas Brownian motion growth behaves
    like t$^2$. A numerical tool is introduced which can be used to
    model dust aggregation in great detail, including the treatment of
    contact forces, aggregate restructuring processes and long-range
    forces.  This tool is used to simulate collisions between magnetic
    grains or clusters and to validate the analytical cross-sections.
    The numerically derived cross section is in excellent agreement
    with the analytical expression.  The numerical tool is also used
    to demonstrate that structural changes in the aggregates during
    collisions can be significant.
\end{abstract}

\section{Introduction}

Coagulation of dust grains is generally believed to be the mechanism
by which growth of particles in the early solar nebula proceeds from
sub-micron grains all the way up to planetesimals
\citep{weid-cuzzi-PPIII,2000prpl.conf..533B}.  However, many questions
remain about the detailed way in which coagulation proceeds, in
particular during the initial phase when the particles are still very
small and dynamically well coupled to the gas.  If the particles are
very strongly coupled to the gas, collisions have to rely on the
relative velocities induced by Brownian motion
\citep{kempf99:_n_partic_simul_dust_growt}.  While these slow
collisions are certain to result in almost perfect sticking between
the grains \citep{Chokshi-ea93,poppe00:_aenal_exper_stick}, it turns
out that they may be too slow  to produce the necessary dust
growth in the limited time frame available
\citep{kempf99:_n_partic_simul_dust_growt}.  First of all, the low
relative velocities severely limit the number of collisions taking
place.  As a second complication, the energies involved in the
collisions are insufficient to change the structure of the aggregates
\citep{coagu,blum00:_exper_stick_restr}.  Simple hit-and-stick growth
in a cluster-cluster aggregation process leads to extremely fluffy
aggregates with low fractal dimensions
\citep{me91,kempf99:_n_partic_simul_dust_growt}.
\citeauthor{kempf99:_n_partic_simul_dust_growt} showed that the
Brownian motion leads to fractal dimensions $D<2$ for which
significant changes in the cross-section-to-mass ratio can only be
expected for very large particles.  The friction time, i.e. the time
in which a particle adapts its velocity to the surrounding gas,
changes in a Brownian growth process only as $r^{1/5}$ with the
particle radius.  Furthermore,
\citeauthor{kempf99:_n_partic_simul_dust_growt} neglected the rotation
of the aggregating particles which will lead to even lower fractal
dimensions of the collision product \citep{bl00}.  Since the
aerodynamic behavior of particles is governed by the friction time
\citep{weid_aerodyn}, the aerodynamic properties of large fluffy
aggregates and of small grains are almost identical.  This poses a
serious problem for the coagulation process which must accelerate
eventually, in order to fulfill the time constraints given.

Possible ways to accelerate growth have been discussed in the
literature, in particular grain settling and radial drift
\citep[e.g.][]{Weid_settling}, or differential coupling to turbulent
eddys of different sizes
\citep{Weidenschilling84,Mizu-eq88,1991A&A...242..286M}.  Furthermore,
aerodynamic concentration is considered as a mechanism to increase
densities locally for enhanced coagulation
\citep{Cuzzi_concentration,1997Icar..128..213K}.  However, most of
these processes rely on a diversity in the aerodynamic properties of
particles: Settling or radially drifting grains will only induce
enhanced collision rates if not all grains settle or drift at the same
rates.  Turbulent eddies will only enhance collision rates if
different grains couple to different eddie scales.
\citet{1997LPI....28.1517W} reports that starting from
  10$^{-4}$\,cm grains, a hight central density of settled dust can
  still be achieved even if the fractal structure of the grains is
  taken into account.

Coagulation may proceed on quite different paths if the basic
mechanism of coagulation is different from what is normally assumed.
An interesting idea was put forward by \citet{nu94} and \citet{nu95}
who studied aggregation of magnetic grains in laboratory experiments.
They found that this accretion process proceeds rapidly forming large
networks of linear aggregates.  \citeauthor{nu94} suggested that such
networks may be important for grain aggregation in the solar nebula.
One way in which this mechanism could change the course of aggregation
is to turn a cluster-cluster aggregation into a particle-cluster
aggregation process.  If the networks of magnetic grains can form
quickly enough, they may eventually provide a large fraction of the
available collisional cross-section in the nebula.  The main
aggregation process for non-magnetic grains would then be the
collision of a grain with such a network.  This mechanism would be
physically similar to a cluster-particle aggregation process which
produces aggregates with fractal dimension $D_f=3$ \citep{Ball84},
leading to more compact structures quickly.

In this article, we will present the foundations for a comprehensive
investigation of magnetic dust aggregation from a theoretical and an
experimental point of view.  We start by reviewing the evidence for
magnetic grains in the solar nebula.  We will then focus on the
physical implications of dipolar interaction between individual dust
grains, i.e. enhanced collisional cross sections and accelerated
aggregation dynamics.  We also present a new numerical tool we devised
for the investigation of dust aggregation including a detailed
treatment of grain-grain interaction through mechanical contacts, and
dipolar (magnetic) forces.  We carried out numerical simulations of
2-body collisions with magnetic grains in order to test the code and
compare the results with the analytically derived cross
sections.  Two subsequent papers will communicate our experimental and
numerical work of magnetic grain growth on the basis of the material
presented here.

\section{Magnetic dust in the solar nebula}

If magnetic interaction is to change the dust aggregation scenario, we
have to assume the existence of a nebular dust component carrying
remanent magnetization \citep{nu94,hn99,hn00a}.  The following
paragraphs present evidence as to the extent this assumption
seems to be justified.

\paragraph{Formation of chondrules}

Many of the small number of pieces that have been found to complete
the puzzle of solar nebular physics come from the study of meteorites.
\citet{jfk88} and \citet{jfk93} provide a comprehensive overview of
this subject.

An important result from the study of meteoritic
material is the occurrence of chondrules, i.e. small spherical
inclusions which are embedded in a presumably more primitive matrix.
Despite a large number of different hypotheses \citep{bo96} as to the
origin of chondrules, the consequences of the necessary melting
process have been studied intensively.  Chondrules are formed when a
physical process in the solar nebula melts silicate dust aggregates
followed by rapid cooling of the molten material which congeals as
spherical droplets \citep{gi97}. Nebula lightning, shock fronts, and
friction have been discussed as possible energy sources. The heating
process itself has been repeatedly investigated both theoretically and
experimentally \citep{le88,gr88}. The heated dust aggregates of
possibly interstellar composition undergo internal reduction-oxidation
reactions, which result in an evaporative loss of metallic iron from
the melt \citep{nu89}.  The iron vapor then condenses into
  sub-micron-sized dust grains.  Depending on the grain size these
  grains can have different magnetic properties
  \citep{nu94,WitheyNuth1999}.  Very small grains below 20\,nm are
  superparamagnetic, large grains $>80$\,nm show multidomain behavior.
  Intermediate sizes, in particular grains between 40 and 60\,nm are
  single domain grains which spontaneously become magnetic dipoles.
In a more general way, melting events in the solar nebula will always
produce magnetic minerals such as magnetite \citep{ri99}.

Chondrules themselves originate from already existing
  aggregates, so the iron grains produced by this process will not have
  been present very early on in order to start the aggregation
  process.  However, similar melting events must have been a
  frequent phenomenon in the early solar system.  The assumption of an
  important population of magnetic iron dust appears plausible on
  these grounds.

\paragraph{The paleomagnetic record in meteorites}

As summarized by \citet{su88}, many meteorites exhibit natural
remanent magnetization (NRM). The carriers of the NRM are magnetic
minerals and metals such as iron-nickel (Fe-Ni: kamacite, taenite),
phyrrotite ($\rm Fe_{1-x}S$), and magnetite (Fe$_3$O$_4$; {\sc
  J\"ackel}, personal communication). A possible source for the NRM in
meteorites are strong magnetic fields that are assumed to have existed
in the solar nebula, magnetizing the dust as it cooled through the
Curie point. 

It is usually assumed that the NMR in meteoritic material was accuired
\emph{after} the formation of the meteorite parent body.  However,
some meteorites show evidence of heterogeneous magnetization, which
could be a hint of magnetization that existed before the dust was
incorporated into the meteorite parent body. The magnetization of
isolated grains is certainly more easily explained than the
magnetization of macroscopic objects, for the latter would require
very stable geometric conditions, i.e., a fixed orientation of the
object that is to be magnetized relative to the magnetizing field.  In
any case, the paleomagnetic record in meteorites also points toward
the existence of a magnetized dust population in the solar nebula.

\paragraph{Cometary dust}

Cometary nuclei are widely supposed to contain the original building
material of our solar system in almost pristine form. The great effort
which is put into prestigious and adventurous space missions like
Rosetta, Stardust, and Deep Space 4/Champollion is a measure
of the importance attributed to the analysis of cometary
matter.

During the Giotto and Vega missions to comet p/Halley in 1986, mass
spectrography of cometary dust grains impacting the spacecraft
gave some insight as to the chemical composition of cometary dust.
\citet{sch97} summarized the results of these
measurements. Accordingly, Halley's dust is composed mainly of
magnesium-rich silicates, but there is an important fraction ($\ge
10$\%) of potentially magnetic minerals and metals similar to those
which have been identified in meteorites, i.e. kamacite
(nickel-iron), phyrrotite ($\rm Fe_{1-x}S$), and magnetite ($\rm
Fe_3O_4$).

Some interplanetary dust particles (IDPs) appear to originate from
comets \citep{1985AREPS..13..147B}.  Up to now, no measurements of the magnetic
properties of IDPs have been obtained (Bradley, priv. comm.).  The
presence or absence of aggregates of magnetic grains in cometary IDP would
be a direct way to determine if a large population of magnetic grains
in the outer solar nebula.

\paragraph{Asteroids}

On its way to the Jovian system, the Galileo spacecraft investigated
the asteroids Ida and Gaspra. The magnetic field measurements in the
vicinity of the two bodies show significant changes in the
solar wind magnetic field as it encounters the asteroids. The magnetic
signatures have been interpreted in terms of a remanent magnetization
of the two asteroids by \citet{ki93}. The same holds for the
  encounter of the Deep Space 1 (DS1) spacecraft with the asteroid Braille in
  1999 \citep{ri01}.

Since asteroids -- which should have sizes similar to
  planetesimals during solar system formation -- are much too small
to sustain a dynamo, their magnetization could be the trace of
primordial magnetic fields or the result of the accretion of
previously magnetized material.

\paragraph{Dust around young stars}
ISO observations of young nearby stars have been used to derive the
dust properties in the disk around stars close to the main
sequence.  Unfortunately, most T Tauri stars are too faint for this,
but their higher-mass counterparts, the Herbig AeBe stars, have proved
to be excellent targets for such a study \cite{HAEBE-review}.  In
several of these stars, the signatures of crystalline silicates have
been found \cite[e.g.][]{malfait-hd100546,Bouwman-ABAur} which are
very likely formed by thermal annealing \citep{bouwman-10um}.  These
silicates are \emph{magnesium-rich}, which may leave the iron
  free to form metallic iron and other potentially magnetic grains.

\section[Magnetic interaction and dust aggregation]{The influence of magnetic interaction on the dust aggregation process}

In this article we focus on magnetic forces as the only
  long-range force.  Electrostatic interactions between charged grains
  could be another source of such long-range interactions between dust
  particles.  Their effect on dust coagulation in a nebular context
  has been investigated both theoretically \citep{ho90,Ossenkopf93}
  and experimentally \citep{2001LPI....32.1262M}.  Since the solar
  nebula is generally considered as optically thick at least in its
  central regions, photoelectric charging can be ruled out when
  discussing possible charging mechanisms for dust grains.
  Triboelectric charging would be due to collisions which usually lead
  to direct sticking within the range of relative velocities
  considered here, even if van der Waals interactions are the only
  attracting force.  Therefore, electrostatic interactions are
  probably of importance only for larger grains
  \citep{2001LPI....32.1262M}.

In the absence of other long-range forces, random thermal agitation is
the sole competitor for magnetic interaction. The relative importance
of these two effects is generally given by a dimensionless parameter
$\lambda$ defined by:
\begin{equation}
  \lambda = \frac{\mu^2\mu_0}{4\pi d^3k_BT}\label{e_mag}\mkomma
\end{equation}
where $\mu$ is the magnetic moment of the particles, $\mu_0$ is the
magnetic field constant and $k_B$ is the Boltzman constant.
The parameter $d$ in Eq.~\ref{e_mag} defines a certain
characteristic interaction distance which is usually taken to be equal
to the particle diameter, $d=2R$, or to the sum of the particle radii,
$d=R_i+R_j$, if we are dealing with multi-disperse grains. For
single-domain particles, which are typically micron-sized or smaller,
$\lambda$ can take on values of the order of $10^5$ at $T=100$ K. In
these cases, the aggregation process will not be diffusive but is
dominated by magnetic forces. Furthermore, magnetic aggregates should
be stable against destructive thermal influences.

According to statistical mechanics, dust particles should acquire a
thermal rotational energy $U_{\rm rot}=k_BT/2$, which amounts to very
high angular velocities for micron-sized particles. In order to have
magnetic interaction overcome this thermal spin-up, the
particle-particle interaction energy must be at least of
the same order of magnitude.  \citet{hn00a} have shown that this
happens inside an interaction radius given by:
\begin{eqnarray}
  R_{\rm m} &=& \bigg( \frac{2\mu_0} {\pi k_BT} \, \mu^2 \bigg) ^
  {1/3}\label{e_rgg}\\[0.2cm] &=& 2d\lambda^{1/3}\nonumber\mkomma
\end{eqnarray}
which is about 10 to 100 times larger than the actual particle
radius for strongly magnetized grains.  It is on these length scales
that magnetized dust particles ``see'' each other and interact. In the
following sections we will examine some of the consequences in more
detail.

\subsection{Collisional cross section}\label{ss_crosssection}

Grain-grain collisions and sticking in the dilute nebular gas result
from relative motion, e.g. due to a thermal distribution of grain
velocities. Owing to the minute geometric dimensions of the dust
particles, the collision probability is usually very low.  Long-ranged
forces, e.g. magnetic forces between magnetized grains, lead to an
enhancement of collisional cross sections.

Usually, the collisional cross section of a (spherical) particle is
simply given by its geometric projection area, i.e.
\begin{equation}
  \sigma = \pi R^2\mkomma
\end{equation}
where $R$ denotes the particle radius.  For two colliding
particles we have:
\begin{equation}
  \tilde{\sigma} = 4\cdot\pi (R_i+R_j)^2\mpkt
\end{equation}
The collision probability in dust aggregation processes depends
directly on the particle cross section: the bigger $\sigma$, the
shorter the mean free time $\overline{\tau}\sim 1/\sigma$ between
subsequent particle collisions.

\citet{hn00a} investigated this
effect for nanometer-sized iron grains numerically, based
on the experimental results obtained by \citet{nu94}.
From scattering simulations with spherical
iron particles with a diameter of 20 nanometers moving at thermal
velocity, $\sqrt{k_BT/m}$, it was shown that:
\begin{equation}\label{e_sigmaenhan}
\frac{\sigma_{\rm m}}{\sigma}(v_{\infty} = v_{th})\approx 900\mpkt
\end{equation}
Since the collision frequency depends on the
cross section, an enhancement as given by Eq.~\ref{e_sigmaenhan}
will cut aggregation time scales by almost three orders of
magnitude.  From Eq.~\ref{e_rgg} we see that the cross section may
be of the same order as the magnetic interaction radius.

In the calculations leading to Eq.~\ref{e_sigmaenhan} we
implicitly assume that magnetized grains are always aligned with the
ambient magnetic field, e.g. an external field or the field created by
other magnetized dust particles. While this should be the case in
terrestrial experiments due to high surrounding gas densities, it is
not obvious that instantaneous relaxation of dipoles should occur in
very dilute media such as the solar nebula.  We will now relax this
assumption.

When a magnetic dipole $\bmu$ interacts with another magnetic dipole
$\bmu'$ over a distance $r$, according to \citet{ja75}
the interaction energy $U$ can be calculated from
\begin{equation}
  U_m = \D\frac{\mu_{0}}{4\pi}\cdot\left[ \frac{\mbox{\boldmath$\mu$}
  \cdot\mbox{\boldmath$\mu$}'}{\D{r^{3}}}-3\frac{(\mbox{\boldmath$\mu$}\cdot\mbox{\boldmath$r$}
  )(\mbox{\boldmath$\mu$}'\cdot\mbox{\boldmath$r$}'
  )}{\D{r^{5}}}\right] \mpkt \label{e_um}
\end{equation}
The most important contributions to the interaction energy come from
distances $r\approx d$ and relative orientations $\theta_{ij}\approx
0$, where $\theta_{ij}$ denotes the angle between two dipoles
contained in a common plane.  Thus, magnetic dipoles tend to align in
a pole-to-pole configuration while minimizing the interaction energy:
\begin{equation}
U_{\rm min} = -\frac{\mu_0}{4\pi}\cdot\frac{2\mu\mu'}{d^3}\;:=
-\frac{\alpha}{d^3}\mpkt\label{e_umin}
\end{equation}
where we have introduced the potential parameter $\alpha$.  An
analytical calculation -- such as the Rutherford formula for
electrostatic interaction -- of the collisional cross section
$\sigma_m$ between magnetized dipolar particles is not possible since
we are not dealing with a central force problem where the interaction
solely depends on the inter-particle distance $r$. In general, the
force between two magnetized dust grains will {\em not} be directed
along the line connecting the particles centers. This is equivalent to
saying that there is no conservation of angular momentum. However,
Eq.~\ref{e_umin} can be used to derive an approximate
expression for $\sigma$ since it essentially describes a
central potential, $U=U(r)$, if we replace the variable $d$ by $r$.
Particle motion in this type of potential will be confined to a plane,
and orbital angular momentum is conserved. The magnetic moments
are assumed to be parallel to a line connecting the particle
centers.

Following the treatment of particle motion in a central potential
\citep[e.g.][]{la84},
scattering of two particles with mass $m$ can be described in terms of
an effective potential
\begin{equation}\label{e_ueff}
U_{\eff}(r)=U(r)+\frac{l^2}{2\overline{m}r^2} \;\mkomma
\end{equation}
in which we have introduced the reduced particle mass, $\overline{m} =
m^2/2m = m/2$ for identical grains. Equation~\ref{e_ueff} combines the
interaction potential $U(r)$ given by Eq.~\ref{e_umin} and the
conservation of orbital angular momentum $l$. We are now dealing with
an equivalent one-body problem.

For scattering problems, $l$ is usually given in terms of the impact
parameter $b$:
\begin{equation}\label{e_l}
  l=\overline{m}\cdot v_{\infty}\cdot b\mkomma
\end{equation}
where $v_{\infty}$ denotes the relative velocity prior to interaction,
e.g. thermal motion. Combining Eqs.~\ref{e_umin} through
\ref{e_l} yields:
\begin{equation}
  U_{\eff} = \frac{\overline{m}\cdot b^{2}\cdot v_{\infty}^{2}}{2\cdot
  r^{2}}-\frac{\alpha}{r^{3}} \mpkt\label{e_ueff2}
\end{equation}
We refer to Eq.~\ref{e_umin} for a definition of $\alpha$. For
potentials with $r^{-2}$-dependence, the maximum impact parameter can
then be derived by simply setting the initial energy
$1/2\,\overline{m}v_{\infty}^2$ equal to the $U_{\eff}$ as given by
Eq.~\ref{e_ueff2}. It should by noted, however, that the effective
interaction potential for dipolar interaction (cf. Eq.~\ref{e_ueff2})
does not display a {\em local minimum} as is the case for
gravitational or electrostatic interaction, but a {\em local maximum}
given by:
\begin{equation}
  U_{0}=\frac{1}{2\alpha^2}\cdot\left(\frac{\overline{m}\cdot
  v_{\infty}^{2}\cdot b^2}{3}\right)^3\mpkt
\end{equation}
We therefore obtain $b_{\rm max}$ by letting $E=U_0$, which implies that
an incoming particle first has to cross a potential wall of height
$U_0$ before it can fall into the center of force:
\begin{equation} 
  \frac{1}{2\alpha^2}\cdot\left(\frac{\overline{m}\cdot
  v_{\infty}^{2}\cdot b_{\rm max}^2}{3}\right)^3 \stackrel{!}{=}
  \frac{\overline{m} v_{\infty}^{2}}{2}\mpkt
\end{equation}
Therefore we find for the maximum impact parameter 
\begin{equation} 
  b_{\rm max} = 3\cdot\left(\frac{\alpha}{\overline{m}
  v_{\infty}^{2}}\right)^{2/3} \mpkt \label{e_bmax}
\end{equation}
and the magnetic cross section is
\begin{equation} 
  \sigma_{\rm m} = 3\pi\cdot\left(\frac{\alpha}{\overline{m}
  v_{\infty}^{2}}\right)^{2/3} \mpkt \label{e_sigma2}
\end{equation}
The central-potential approximation necessarily overestimates the
magnetic cross section by a factor or order 2, because only
maximum interaction is taken into account when deriving
Eq.~\ref{e_sigma2}.

\subsection{Accretional remanence}\label{ss_accrem}

A direct consequence of the increased particle cross section for
magnetic grains is the selective coagulation of magnetized dust
particles with their kin. Undisturbed magnetic aggregation should lead
to the formation of chain-like configurations with very low fractal
dimension, since magnetic grains tend to coagulate in a pole-to-pole
pattern. $D_f=1$ characterizes a perfect chain or rod. With
non-magnetic dust grains, aggregation processes in three-dimensional
Euclidian space usually result in structures with fractal dimensions
around 2
\citep{ke97,wurm97:_exper_prepl_dust_coagul_aggreg,wu97,Weid-ea88}. More
recently however, \citet{bl00} and 
\citet{blum00:_exper_stick_restr} reported on the growth of
aggregates of $\rm SiO_2$-spheres with elongated structures and low
fractal dimension ($D_f=1.3$ \citep{bl00}) which can be attributed to
rotational effects.

In the case of magnetic grains, chain-like aggregation should coincide
with an efficient conservation of the total magnetic moment. For
perfectly aligned dipolar particles, the magnetic moment of a chain is
given by the sum over the individual magnetic moments $\mu_i=\mu$:
\begin{equation}
  \mu_{\chain}=\sum\limits_{i=1}^N \mu_i\;=N\cdot\mu\mpkt
\end{equation}
For sufficiently high values of the magnetic interaction parameter
$\lambda$ (cf. Eq.~\ref{e_mag}), i.e. $\lambda\gg 1$, there will
be no statistical mixing of individual magnetic moments, but --
leaving aside thermal agitation and locally disturbed agglomeration
due to multi-particle interaction -- an increase of the magnetic
moment per cluster with the number of constituent particles.  This
increase $\mu_{\agg}=\mu_{\agg}(N)$ is called {\em accretional
  remanence} \citep{hn99,hn00a}.  Its functional dependence was
determined from numerical simulations \citep{hn00a} to be
\begin{equation}
  \mu_{\agg}(\mu , N) = \mu\cdot N^{0.63}\mkomma
\end{equation}
where $\mu=\mu_i$ again denotes the magnetic (dipole) moment of a
single grain. Large aggregates assembled from thousands of small
magnetized grains will therefore still show a large fraction of the
total magnetic moment of their constituents.

\begin{figure}[h]
  \begin{center}
    \includegraphics[width=\textwidth,clip=]{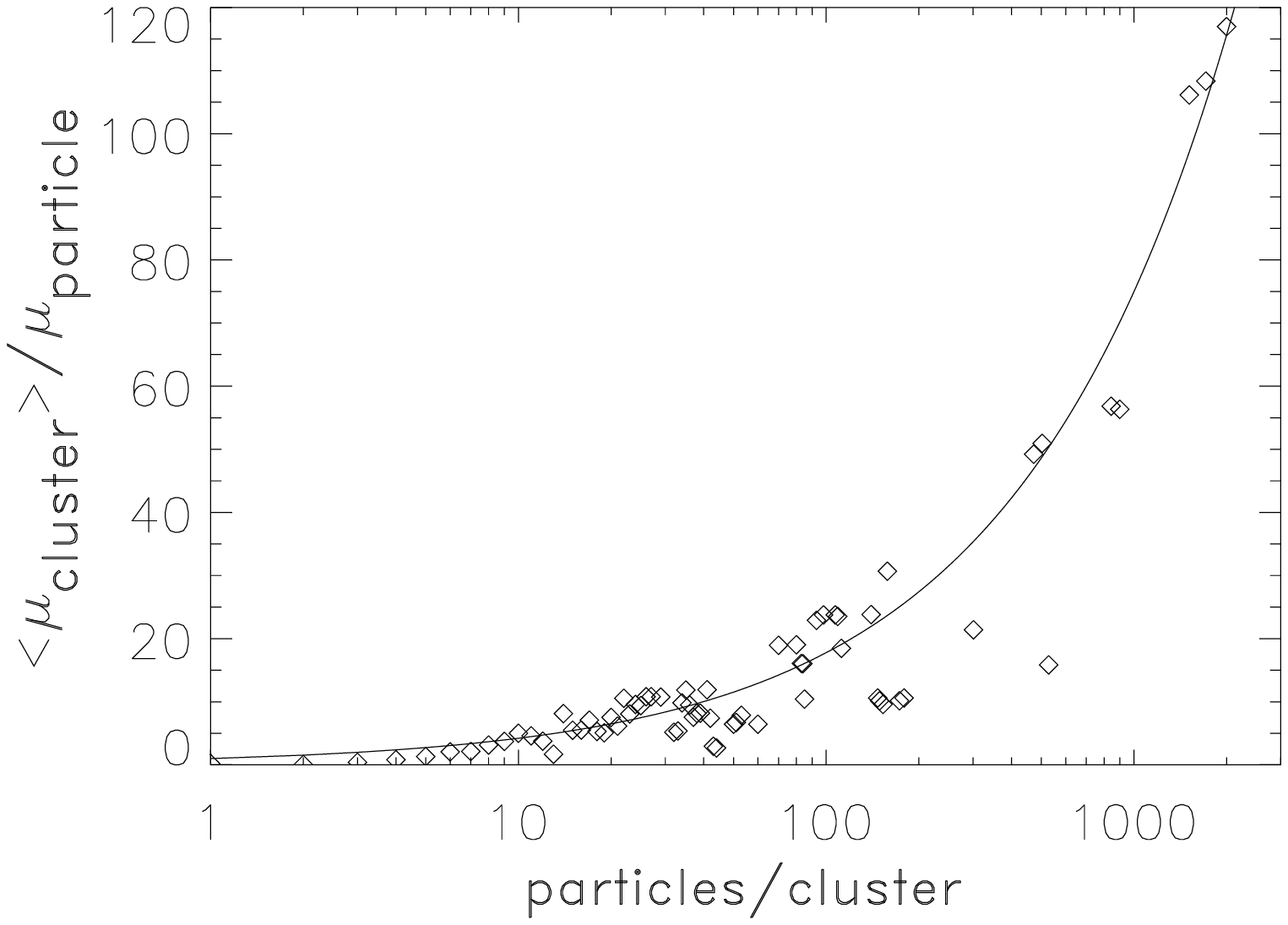}
    \caption[Accretional remanence.]{\label{f_mags}{\em Accretional
    remanence}: The increase of cluster magnetic moment with the
    number of individual magnetic grains in numerical simulations
    \citep{hn99,hn00a}. The curve illustrates the
    relationship $\mu\propto N^{0.63}$, which shows the strong
    tendency of magnetized grains to aquire a state of minimum
    magnetostatic energy (cf. Eq.~\ref{e_umin}), i.e. a
    pole-to-pole configuration for each pair of grains, and the
    misalignment effect caused by thermal agitation as well as the
    microphysics of the coagulation process.}
  \end{center}
\end{figure}
 
\subsection{Aggregation Dynamics}

Due to its irreversible character, studies of the the kinetic
properties of particle aggregation focus on the dynamic properties of
the distribution of clusters. Within this context, the features of
aggregation are described in terms of the cluster-size distribution
$n_s(t)$ as a function of time \citep[e.g.][and references
therein]{mi99a}. As is the case for all distribution functions, $n_s$
can be used to calculate higher moments, in particular the mean
cluster size $S(t)$ which is defined by:
\begin{equation}\label{e_st}
  S(t) = \frac{\sum\limits_ss^2n_s(t)}{\sum\limits_ssn_s(t)}\mkomma
\end{equation}
the variable $s$ being one particular cluster size from the
distribution. The asymptotic behavior of the mean cluster size is a
power law
\begin{equation}
  S(t)\sim t^z\label{e_dynco}
\end{equation}
with the so-called {\em dynamic exponent}, $z$, which depends on the
aggregation process. For ensembles composed of identical particles,
the mean cluster size can be identified with the mean aggregate mass
$m(t)$.

To our knowledge, the dynamical evolution of magnetic aggregation has
never been analyzed before in an astrophysical context, i.e. involving
micron-sized particles, low number densities, and a regime of free
molecular flow\footnote{In the regime of free molecular flow, the mean
free path of the molecules of the embedding medium (gas) is greater
than the size of the dust particles under consideration.}. According
to \citet{bl00}, we expect a behavior (cf. Eq.~\ref{e_dynco}):
\begin{equation}
  m(t) \propto t^z
\end{equation}
for a quasi-monodisperse system. For an ensemble of aggregating
particles, the dynamic exponent $z$ depends on the evolution of the
collisional cross section $\sigma\propto m^{\eta}$ and of the
collision velocity $v_c \propto m^{\nu}$. The exponent $z$ can be
derived using the following equation \citep{bl00,me91}:
\begin{equation}\label{e_etanu}
  \eta+\nu = \frac{z-1}{z}\mpkt
\end{equation}
Just like $m(t)$, $\sigma_c$ and $v_c$ are assumed to obey power
laws. Common dust aggregates have $D_f=2$, therefore $\sigma_c\propto
m$ and $\eta = 1$. Thermally induced relative velocities scale with
$m^{-1/2}$, i.e. $\nu=-1/2$. For such
systems, we expect to find a dynamic exponent $z=2$.

In magnetically driven coagulation in free molecular flow, particle
motion is not diffusive but governed by strong inter-particle
forces. For aligned grains with identical dipole moments $\mu$ we thus
have (cf. Eq.~\ref{e_umin}):
\begin{equation}\label{e_vc}
  v_c \propto \frac{\mu (m)}{\,\sqrt{m}\,}\mpkt
\end{equation}
In Eq.~\ref{e_vc} we have to insert the functional dependence $\mu
(m)$ of the aggregate dipole moment on the aggregate mass. As shown in
the previous section, this dependence has been found in numerical
simulations \citep{hn00a,hn98} to be $\mu (N) = \mu (m) \propto
m^{\gamma}$ with $\gamma = 0.63$. In combination with Eq.~\ref{e_vc}
we obtain $v_c\propto m^{0.63}\cdot m^{-1/2} = m^{0.13}$, i.e. $\nu
\approx 0.13$.

In contrast to diffusion-limited aggregation, the collisional cross
section of magnetic grains is greatly enhanced over the geometric
particle size. Again, for perfectly aligned dipoles we have:
\begin{eqnarray}
  \sigma_c & \propto & \left( m^{\gamma}
  \right)^{4/3}\mkomma\nonumber\\[0.1cm] &=& \left( m^{0.63}
  \right)^{4/3}\mkomma
\end{eqnarray}
thence $\eta \approx 0.84$.

Since the foregoing derivations of $\nu$ and $\eta$ imply maximum
magnetic interaction, these values have to be considered as strict
upper limits. To treat this problem in a more quantitative manner, we shall include the increasing dipole
relaxation time $\tau (m)$, i.e. the characteristic mass-dependent
time scale for dipole alignment.

The relaxation time $\tau$ associated with a rotation of dipoles,
e.g. through an angle $\varphi = \pi / 2$, can be estimated from:
\begin{equation}
  \tau \propto
  \sqrt{\,\frac{\varphi}{\dot{\omega}}\,}\;\,\sim\;\,\sqrt{\,\frac{I}{M}\,}\;\,\sim\,\;\sqrt{\,\frac{m^2}{\mu
  (m)^2}\,}\mkomma
\end{equation}
where $\dot{\omega}$, $I$, and $M$ denote the magnetic angular acceleration, the moment of inertia of the aggregate, and the magnetic
inter-particle torque, respectively. Thus we obtain $\tau \propto
\sqrt{m^{2\cdot (1-0.63)}} = m^{0.37}$. If we include this aspect in
our calculation of the exponents $\nu$ and $\eta$, we find the
alignment-corrected values $\nu^{\star}$ and $\eta^{\star}$ by
replacing $\mu^2$ in Eq.~\ref{e_umin} with
$\mu^{2-0.37}= \mu^{1.63}$ to account for imperfect alignment. The
results are:
\begin{eqnarray}
  \nu^{\star} & \approx & 0.01\label{e_nustar}\mkomma\\[0.1cm]
  \eta^{\star} & \approx & 0.68\label{e_etastar}\mpkt
\end{eqnarray}

Insertion of $\eta^{\star}$ and $\nu^{\star}$ into Eq.~\ref{e_etanu}
yields $z\approx 3.2$. This value for the dynamic exponent should
characterize the kinetics of magnetic dust aggregation. It is the
direct consequence of the combined effects of enhanced collisional
cross sections and accretional remanence.

\section{Simulations using \sand: Two test cases}
\label{testcases}

A thorough analysis of the intricate microphysics of magnetic
aggregation requires numerical tools which are able to treat both
long-range magnetostatic and short-range surface forces for large
particle systems. To this end, we devised a new coagulation code
called \sand.  Please refer to Sect.~\ref{s_sand} for a detailed
description of the program which was also used for additional
numerical simulation of magnetic coagulation. The results of this work
will be the subject of a subsequent paper. In this section, we present
two test cases which are meant to illustrate the abilities of \sand.

\subsection{Cross sections}

To test the validity of our theoretical treatment of magnetic
aggregation, we performed detailed one-on-one scattering simulations,
in which one dipole is magnetically deflected or attracted in the
field of an identical, initially motionless, counterpart.

For a detailed study of magnetic cross sections, we systematically
varied the impact parameter around the value $b_{\rm max}$ as given by
Eq.~\ref{e_bmax}.  We are thus able to determine the maximum
value for collisions at different initial relative velocities
$v_{\infty}$, which were taken from a Maxwellian distribution at 100 K
($0.3\cdot v_{th}\le v_{\infty}\le 3\cdot v_{th}$).

\begin{figure}[p]
\includegraphics[width=\textwidth,clip=]{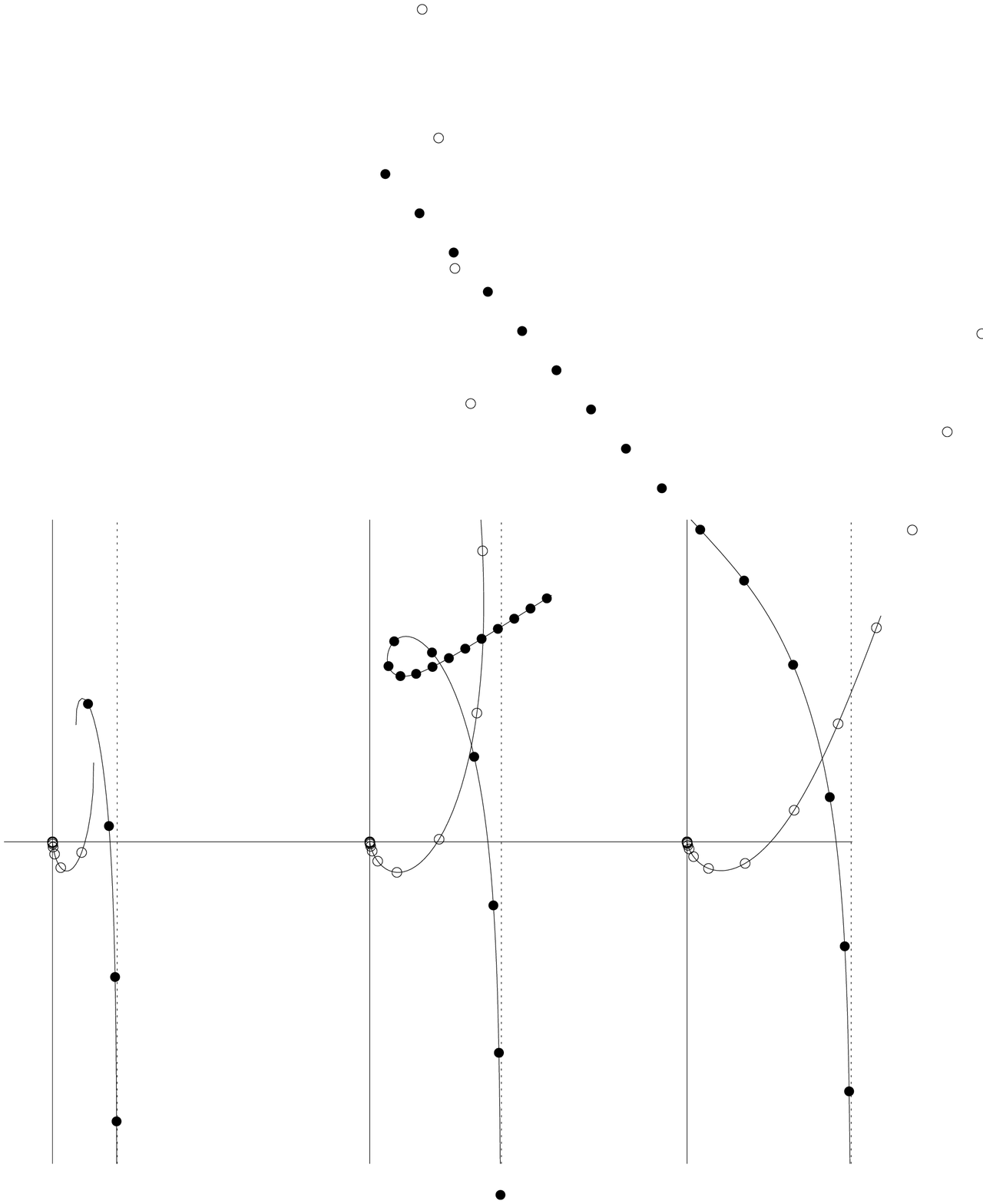}
\caption{\label{fig:trajectories} Three example trajectories of
  numerical collision experiments with \sand.  The particles are
  20\,nm iron grains with a magnetic moment of $200\, {\rm A}{\rm
    m}^2{\rm kg}^{-1}$ and thermal rotations derived from a Maxwellian
  distribution at 100\,K.  Initial velocity is $v_{\mathrm{th}}/16$.
  The impact parameters are 0.95, 1.0, and 1.025 times the
  analytically derived maximum impact parameter $b_{\rm max}$,
  respectively.  Note that the size of the dots does not indicate the
  size of the grains which would be hardly visible on this scale since
  $b_{\rm max}\approx 250 R$.}
\end{figure}

\begin{figure}[p]
       \subfigure[Low velocity part.]
      {\includegraphics[width=8cm]{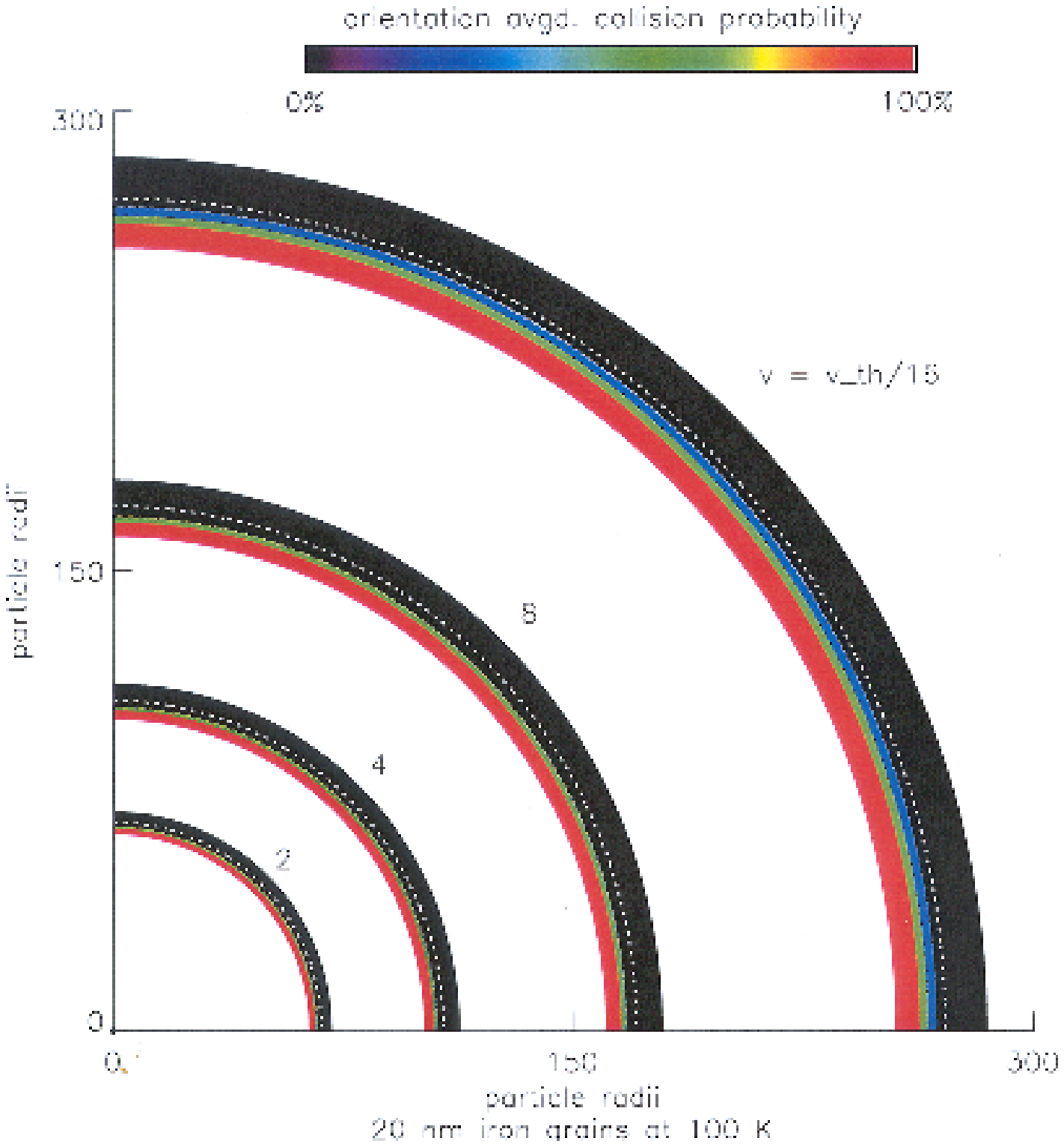}}
      \hfill
        \subfigure[High velocity part.]
      {\includegraphics[width=8cm]{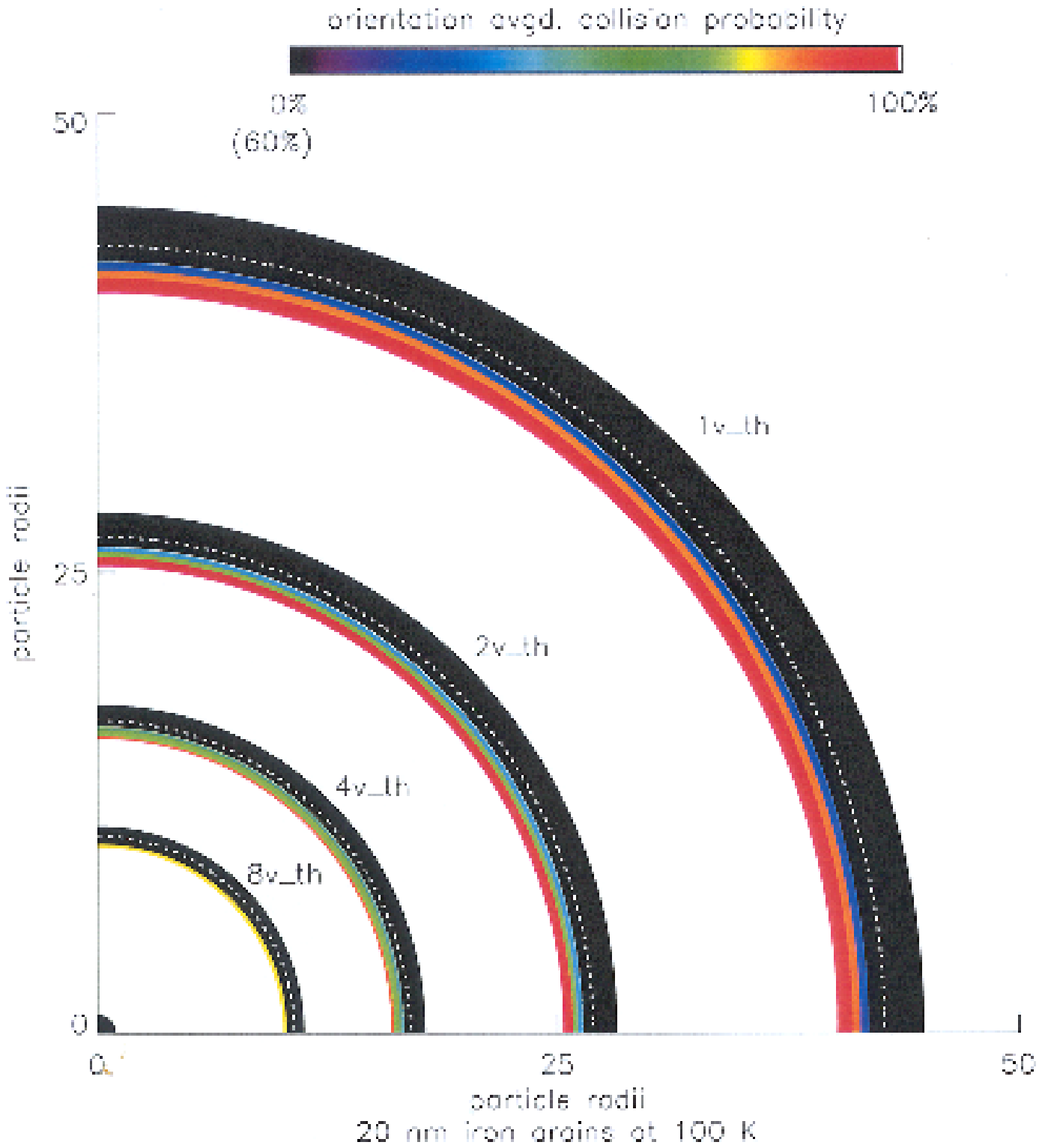}}
     \caption[Magnetic cross section II.]{\label{f_scatt2} Numerical
       validation of Eq.~\ref{e_sigma2}.  Collision probabilities for
       different relative velocities and impact parameters are
       orientation averaged and color-coded. {\em Black} means that
       for no relative orientations of dipoles a collision could be
       detected. {\em White} means that all randomly chosen
       orientations resulted in a collision. For one specific
       velocity, all the area {\em outside} the black annular region
       should be black as well, the same holds for the area {\em
         inside} the white annular region, which should be all white.
       Please note that the color bar has been compressed for the high
       velocity regime, with black coding meaning $p\le 60$\%. The
       dashed lines denote the theoretically derived cross section
       according to Eq.~\ref{e_sigma2}.}
\end{figure}

For all configurations $[v_{\infty}\mkomma b]$ we randomly chose the
initial orientations ($\theta_i\mkomma \phi_i$) of the two scattering
dipoles, which were placed at a relative distance as given by
Eq.~\ref{e_rgg}.  Figure~\ref{fig:trajectories} shows three example
trajectories.  Only the first scattering calculation with an impact
parameter $b=0.95\cdot b_{\rm max}$ leads to a ``hard'' collision of
the grains.  The second calculation with an impact parameter $b=b_{\rm
  max}$ does not lead to a ``hard'' collision.  Instead we see that
the two grains almost entirely exchange their energy and momentum.
The impacting grain is moving only very slowly after the collision,
while the grain initially at rest leaves the scene at a speed
comparable to the impactor's initial velocity.  In
Fig.~\ref{fig:trajectories}, different velocities are represented by
different spacing of dots along the trajectory.  For the third
calculation with $b=1.05\cdot b_{\rm max}$, the grains do not hit each
other and there is only partial transfer of energy.  These three
calculations therefore show good qualitative agreement between the
predicted cross section (c.f. Eq.~\ref{e_sigma2}) and the numerical
result.

Figures \ref{f_scatt2}(a,b) analyze the results of a large number of
simulations in a quantitative way, using color coding to account for
orientation dependent collision probabilities.  While the detailed
values are generally smaller than those obtained previously
\citep{hn99,hn00a}, the qualitative agreement is good. At
$v_{th}=\sqrt{k_BT/m}$ we find:
\begin{equation}\label{e_sigmaenhan2}
\frac{\sigma_{\rm m}}{\sigma}(v_{\infty} = v_{th})\approx 130\mpkt
\end{equation}

The discrepancy between the two results presented in
Eq.~\ref{e_sigmaenhan} and Eq.~\ref{e_sigmaenhan2} can be interpreted
as a direct measure of the effect of the degree of magnetic
alignment, although it should be noted that a small part of the
discrepancy is due
to the change in temperature (80 K vs. 100 K).  According to
Eq.~\ref{e_sigma2}, $\sigma_m$ scales with $T^{-2/3}$. The
temperature-corrected value for $\sigma_m$ in Eq.~\ref{e_sigmaenhan2}
should therefore be approximately $130\cdot 0.8^{-2/3}\approx 151$.
The numerical values equally validate Eq.~\ref{e_sigma2}. As had to be
expected, the theoretically derived value for $\sigma_m$ -- the dashed
line in Fig.~\ref{f_scatt2} -- is too big, since only maximum
interaction enters the central-force approximation.  Orientation
averaged collision probabilities reach 100\% at smaller impact
parameters that can be estimated from the outer radius of the white
annuli, $R_w$.  This can be accounted for by introducing an empirical
correction factor $c=(R_w/R_{dash})^2$ in Eq.~\ref{e_sigma2}:
\begin{equation}
  \sigma_{\rm m} = c\cdot 3\pi\,\left(\frac{\alpha}{\overline{m}
  v_{\infty}^{2}}\right)^{2/3} \mpkt \label{e_sigma3}
\end{equation}
By inspection of Fig.~\ref{f_scatt2} we estimate $c$ to lie between
0.8 and 0.9, i.e. the approximation error in Eq.~\ref{e_sigma2}
is of the order of 10-20\%. We do not see a strong velocity dependence
of this correction factor.

\begin{figure}[p]
\includegraphics[width=\textwidth]{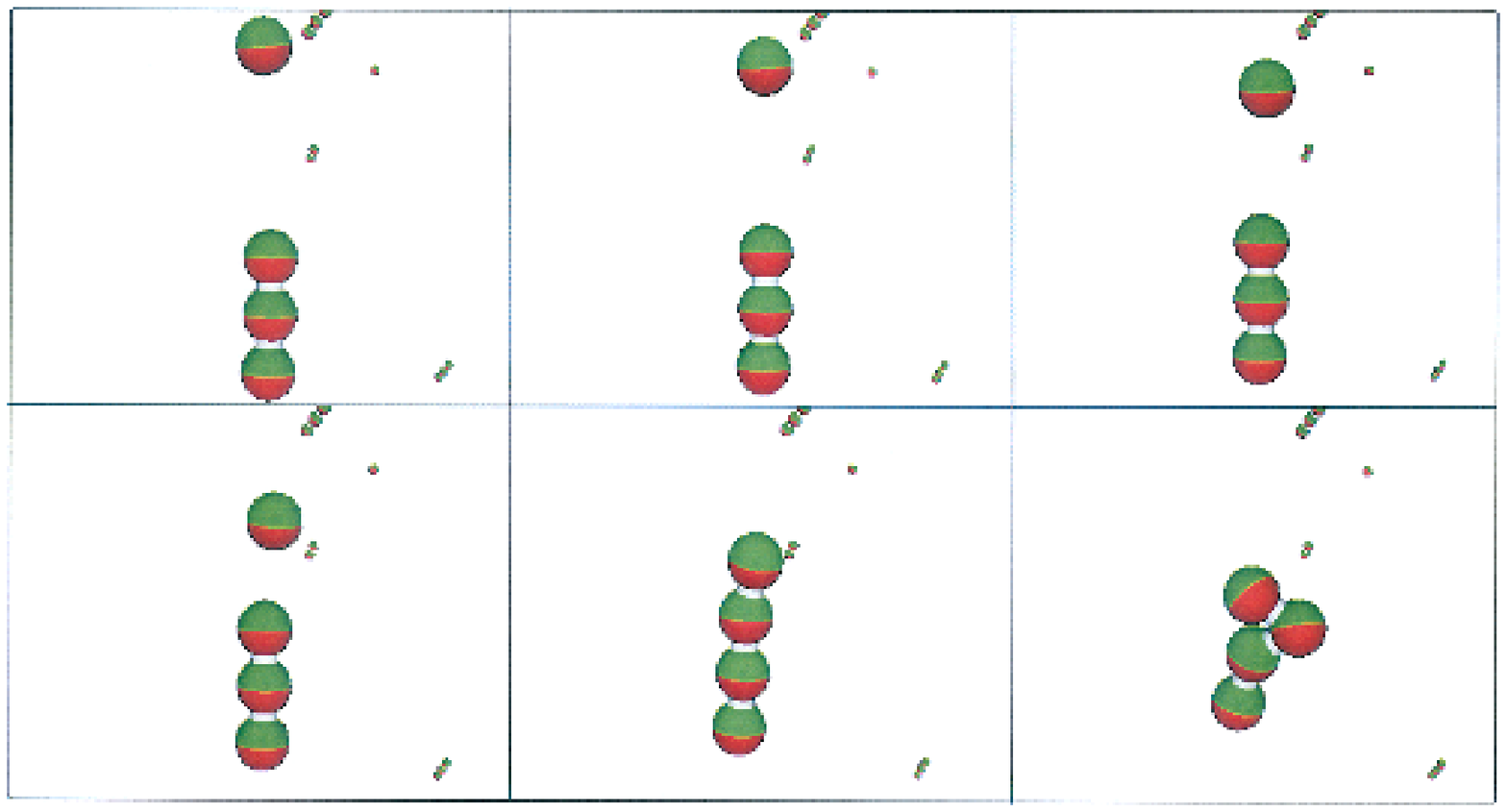}
\caption{\label{fig:bending} Example of a collision between a chain and
  a single particle which illustrates that the detailed treatment of
  contact forces is important to determine the correct structure
  of an aggregate.  The impact of the incoming grain moves the grain
  at the end of the chain.  The resulting T-like structure would not
  have resulted from a simple hit-and-stick approximation.}
\end{figure}

\subsection{Restructuring and contact forces}

The long-range interactions between magnetized dust grains give rise to
an increase in inter-particle collision velocities which have to be
accounted for in realistic coagulation models.
Figure~\ref{fig:bending} illustrates the importance of including
configurational flexibility and means for aggregate restructuring in
coagulation models.
Model calculations of dust coagulation usually dwell on very few
assumptions, sometimes neglecting the aggregate structure altogether.
The simplest approach treats grain growth in analogy to the
coagulation of liquid drops, where the combination of two drops leads
to the formation of one -- bigger -- drop. Next in complexity are
``hit-and-stick'' models \citep[e.g.][]{Ruzmai,Ossenkopf93,MeakinDonn88} where
grains are assumed to stick where they hit, whence the name, and do
not influence the aggregate structure. While the particle trajectories
may be either ballistic \citep[e.g.][]{Ossenkopf93}) or stochastical,
i.e. pseudo-ballistic with random changes of direction on a
diffusion-controlled time scale \citep{ke97}, there is no aggregate
restructuring during collisions. Some models include sticking
probabilities \citep{Ossenkopf93} whereas others prescribe fixed
values for the fractal dimension at different stages of the simulation
\citep{Weid-ea88}.  A realistic treatment of aggregates as solid
bodies has been given by \cite{richardson95}.

The detailed analysis of inter-particle contact forces by
\citet{coagu} allows a more consistent approach. Clusters of particles
are no longer treated as separate objects that must be parameterized
with respect to their physical properties, rather they
preserve their identity and are considered under the influence of
all forces transmitted through each grain's contacts with
adjacent particles. Especially when strong inter-particle
forces are involved, grain-grain collisions can be violent and produce
intricate results.  This is shown in Fig.~\ref{fig:bending} where the
linear aggregate is restructured to a T-shape after the fast impact of
another grain.

\section{Conclusions}

The material presented in this paper lays the foundation for a
detailed analysis of the effects of magnetic aggregation for the
accumulation of solid matter in the early solar system.
Generalizing previous work by \cite{hn99,hn00a}, we have derived an
analytical expression for the collisional (capture) cross section of
magnetic dust grains. As pairs of dipolar magnetic particles tend to
line up in pole-to-pole configuration, aggregates of such particles
``preserve'' the magnetic moment of their constituents, thus
displaying ``accretional remanence''. The combination of enhanced
cross section and accretional remanence is responsible for accelerated
growth dynamics.  This fact is quantitatively described by a dynamic
exponent, $z$, such that the temporal dependence of the mean aggregate
mass, $m$, takes on the form of a power law, i.e., $m\propto t^z$.  We
have shown that the dynamic exponent takes on a value $z \approx 3.2$ for
magnetic aggregation, whereas $z = 2$ for aggregation processes with
non-magnetic particles. 

Part of our theoretical approach has been validated numerically as a
means for introducing a new coagulation code which we devised for the
detailed analysis of magnetic coagulation. Its architecture allows the
study of systems of particles interacting via long-range magnetostatic
coupling as well as through short-range elastic forces. Aggregate
restructuring is included. We provide evidence that such an
approach -- although computationally rather costly - is required for
physically realistic modelling.  A forthcoming paper will present
extensive numerical studies of magnetic aggregation.

{\bf Acknowledgements:} We would like to thank Tijmen van de Kamp for
the development of Grainview, an interactive tool to visualize
aggregate simulations; Vincent Icke and Kees Dullemond for discussons
about how to implement grain rotation, Josef Sch\"ule for help with
optimizing the code.  CD acknowledges financial support from Pioneer
grant 600-78-333.  HN acknowledges financial support from the
Studienstiftung des Deutschen Volkes and from DFG grant GL~142/13-1.

\appendix
\section{The \sand code}
\label{s_sand}

\sand (Soft Aggregate Numerical Dynamics) is a code to compute the
aggregation of dust particles in a detailed way, treating the
microphysics of the contacts between dust particles, long range forces
like magnetic dipole and electrostatic interactions, and external
forces like external magnetic or gravitational fields.  This is a new
combination which in this way has not been used before. The
precursor of \sand was the 2D code used by \citet{coagu} to study the
behavior of dust aggregates in collisions, but without long-range
forces. 

\sand treats aggregates of dust grains as N-particle systems.  Each
individual grain is free to move under the forces acting on it.  Such
forces can be due to external fields, to long-range grain-grain
interactions, and to mechanical force transmission through physical
contacts between grains.

\sand currently treats only spherical particles, but an extension to
sphe\-ro\-ids or ellipsoids would be possible.

\sand was original written to treat contact forces only, and therefore
was best suited to study the collision of aggregates.  Full
aggregation calculations were not possible because the detailed
numerical treatment of the contacts imposes very short time steps, of
the order of 10$^{-11}$ seconds in order to resolve the vibrational
motion of grains in contact.  Long-range forces considerably shorten
the growth-time for an aggregating particle system.  It therefore
becomes feasible to conduct coagulation calculations with the code.  On
the other hand, implementing long-range forces also brings up the
following challenges.

\begin{itemize}
\item[$\bullet$] For long-range forces, every particle in an
  $N$-particle numerical simulation interacts with {\em all other
    particles}, thus leading to an $N^2$ scaling for the computation
  time.  If only contact forces are involved, the code is linear
  in $N$.
\item[$\bullet$] Some kind of boundary condition has to be introduced
  in order to model the interaction of ``edge particles'', i.e.
  particles close to the edge of the
  spatially limited simulation volume. For forces of virtually
  unlimited range, this can be a serious complication.
\item[$\bullet$] As interaction forces are strongly dependent on the
  inter-particle distance, interaction between grains that are far
  apart will be treated in too much detail -- numerically
  speaking -- if only one global simulation time step is used. This
  problem is of particular importance in the case of \sand,
  since strong contact forces require very small time steps in the
  numerical treatment of the differential equations involved.
\end{itemize}

\sand tackles these problems by means of a list-based interaction
calculation for long-range forces and a multiple time step algorithm,
that separates short-range contact forces from long-range dipolar
interaction.

In the following we describe briefly the equations solved by \sand.
Two test calculations are discussed in Sect.~\ref{testcases}.
More detailed simulations of magnetic accretion will be the subject of
a forthcoming paper.

\subsection{Grain properties}

Each individual grain $i$ is described by material properties
(specific density $\rho_i$, surface energy $\gamma_i$), radius $a_i$,
mass $m_i$, moment of inertia $I_i$, magnetic moment vector
$\vec{\mu}_{i}$.  The state of each particle is given by
a location $\vec{r}_i$, a velocity $\vec{v}_i$, rotational state
$\roti$, and by a rotation velocity $\vec{\omega}_i$.

\subsection{Linear Motion}

The differential equations describing the linear motion of each grain
are simply
\begin{align}
\label{eq:4}
\dot{\vec{r}_i} &= \vec{v}_i \\
\dot{\vec{v}_i} &= \vec{F}_{i,\rm contact} + \vec{F}_{i,\rm magnetic} \mpkt
\end{align}
$F_{i,\rm contact}$ is the sum of all linear elastic forces
transmitted through contacts involving grain $i$.  $F_{i,\rm
  magnetic}$ is the sum of all magnetic interactions of grain $i$ with
other grains and an external magnetic field.  In a similar way,
gravitation and electrostatic forces could be added, but this is not
relevant for our current study.

\subsection{Rotation}

The rotational state $\roti$ is a four-vector which contains the
rotation parameters $\rotix$, $\rotiy$, $\rotiz$, and $\rotit$ which
can be used to describe the rotation of a solid body by differential
equations, avoiding the singularities characteristic for Eulerian
angles and similar 3-parameter representations.  We follow the
formalism by \citet{Whittaker}.

Grains initially enter the calculations with parameters $\roti(t_0)
= (0,0,0,1)$.  At a later state, $\roti(t) =
(\rotix(t),\rotiy(t),\rotiz(t),\rotit(t))$.  The new orientation can
be computed from the old orientation by a single rotation using the
rotation matrix

\begin{equation}
\label{eq:1}
\rotmat_i(t) = 
\begin{pmatrix}
\begin{array}{r}
 \rotix\rotix - \rotiy\rotiy  - \\  \rotiz\rotiz + \rotit\rotit 
\end{array} &
2(\rotix\rotiy - \rotiz\rotit) &
 2(\rotix\rotiz + \rotiy\rotit) \\[4mm]
 2(\rotix\rotiy + \rotiz\rotit) &
\begin{array}{r}
 -\rotix\rotix + \rotiy\rotiy - \\  \rotiz\rotiz + \rotit\rotit
\end{array} &
 2(\rotiy\rotiz - \rotix\rotit) \\[4mm]
 2(\rotix\rotiz - \rotiy\rotit) &
 2(\rotiy\rotiz + \rotix\rotit) &
\begin{array}{r}
 -\rotix\rotix - \rotiy\rotiy + \\  \rotiz\rotiz + \rotit\rotit
\end{array}
\end{pmatrix} \mkomma
\end{equation}
where we have omitted the time arguments for brevity.  This matrix can
be used to rotate any vector properties of the particle, such
as the magnetic moment into the current state.

The differential equations for the rotation parameters are given by
\begin{align}
\label{eq:2}
\rotixdot  &= \;\;\; ( \rotit\ox + \rotiz\oy - \rotiy\oz)/2 \nonumber \\
\rotiydot  &= \;\;\; ( \rotit\oy - \rotiz\ox + \rotix\oz)/2 \nonumber \\
\rotizdot  &= \;\;\; ( \rotiy\ox - \rotix\oy + \rotit\oz)/2 \nonumber \\
\rotitdot  &= -(\rotiz\oz + \rotiy\oy + \rotix\ox)/2 \mpkt
\end{align}
The rotation parameters also must fulfill a normalization condition, i.e.
\begin{equation}
\label{eq:5}
\rotix^2+\rotiy^2+\rotiz^2+\rotit^2=1 \mpkt
\end{equation}
Solving the differential Eq.~\ref{eq:2} introduces
numerical errors into the rotation parameters.  After each successful
timestep, we therefore re-normalize the rotation parameters in
accordance with Eq.~\ref{eq:5}.

The differential equation for the change of the angular velocity
$\vec{\omega}$ is simply
\begin{equation}
\label{eq:3}
\dot{\vec{\omega}_i} = \vec{M}_{i,\rm contact} + \vec{M}_{i,\rm
  magnetic} \mkomma
\end{equation}
where $\vec{M}_{i,\rm contact}$ is the sum of all elastic torques
transmitted through contacts involving grain $i$.  $\vec{M}_{i,\rm
  magnetic}$ is the torque due to all magnetic interactions of the
grain $i$ with other grains and with an external magnetic field.

\subsection{Magnetic forces}

The basic equation for the magnetostatic interaction between two
dipolar particles has been given in Sect.~\ref{ss_crosssection}
(cf. Eq.~\ref{e_um}). For notational simplicity, we let the
constant factor $(\mu_0/4\pi ) = 10^{-7}$ be equal to 1 for the
remainder of this section. The derivation of the force acting between
two magnetic dipoles $i,j$ is straightforward. In three-dimensional
Cartesian coordinates, we have:
\begin{equation}
  \bF_{ij} = \left( \bmu_j\bnabla\right)
  \bB_{ij}\label{e_dipdip1}\mpkt
\end{equation}
In Eq.~\ref{e_dipdip1}, the double subscript $ij$ describes
the influence of particle $i$ on particle $j$, i.e. at the location of
particle $j$. Individual particles are referred to by single
subscripts, e.g. $i$.

Carrying out the multiple differentiations in Eq.~\ref{e_dipdip1}
and introducing the relative distance vector, $\br_{ij}$, of length
$r$ we have:
\begin{eqnarray}
  \bF_{ij} &=& \frac{3}{r^5}\left[\;\mat{M}_1 - \frac{5\left(\bmu_i\cdot \br_{ij}\right)}{r^2}\cdot\mat{M}_2\;\right]\cdot\bmu_j\label{e_fmag}\mkomma\\[0.1cm]
    \mbox{with}\;\;\;\;\;\mat{M}_1 &=& \bmu_i\cdot\br_{ij}{}^T + \br_{ij}\cdot \bmu_i{}^T + \left(\bmu_i\cdot \br_{ij}\right)\cdot\mat{I}\label{e_m1}\mkomma\\[0.1cm]
    \mbox{and}\;\;\;\;\;\;\mat{M}_2  &=& r_{ij}\cdot r_{ij}{}^T\mpkt\label{e_m2}
\end{eqnarray}
In Eqs.~\ref{e_m1} and \ref{e_m2}, the superscript $T$ denotes a row
vector. The matrix $\mat{I}$ in Eq.~\ref{e_m1} is the identity matrix.
All matrices contained in Eqs.~\ref{e_fmag}-\ref{e_m2} are symmetric,
i.e., for actual computation only six out of nine elements have to be
calculated. Trivially, $\bF_{ji}=-\bF_{ij}$.

Apart from translational forces, dipoles will also exert torques
$\bM_{\rm int}$ on their counterparts. Each contribution
$\bM_{ij,\rm int}$, i.e. the torque acting on dipole $j$ in the
magnetic (dipole) field due to particle $i$ at the location of $j$, is
given by
\begin{eqnarray}
  \bM_{ij,{\rm int}} &=& \bmu_j\times\bB_{ij}\label{e_mint1}\\[0.1cm]
  &=& \bmu_j\times\left[\; 3\;\frac{\br_{ij}\cdot\left( \bmu_i\cdot\br_{ij}  \right)}{r^5} - \frac{\bmu_i}{r^3} \;\right]\label{e_mint2}\mpkt
\end{eqnarray}

In addition to mutual interaction (cf. Eq.~\ref{e_mint2}),
magnetic grains can interact with an external magnetic field
$\bB_{\rm ext}$. As a result of an undamped interaction the perpendicular
part of the magnetic dipole moment vector of each grain,
$\mu_j^{\star}=\bmu_{j,\bot\bBT}$, will oscillate around the
direction of the external field. The magnetic torque
$\bM_{j,{\rm ext}}$ causing this oscillation is given by
\begin{eqnarray}
  \bM_{j,{\rm ext}} &=& \bmu_j\times\bB_{\rm ext}\mkomma\\[0.1cm]
  \left|\bM_{j,{\rm ext}}\right| &=& \mu_j^{\star}\cdot B_{\rm ext}\mpkt
\end{eqnarray}
In the scope of \sand, the external magnetic field will be
constant in time and -- without loss of generality -- directed along
the $z$-direction.

\subsection{Contact forces}

When two grains collide, they can form a contact.  Forces are
transmitted through such contacts, and the magnitude of these forces
largely determines the structure and mechanical properties of an
aggregate.  A contact between two grains has an equilibrium state in
which the attractive forces are balanced by elastic forces due to
compression of the material near the contact
\citep{rolling,sliding,coagu}.  Dynamic forces on the grains lead to
additional stresses in the contact area which are responsible for
contact forces acting on the two grains.  As long as the forces do not
exceed the elastic regime, the contact forces can be calculated as a
function of the displacement of the grains from the equilibrium
contact position.  Therefore, the problem of computing contact forces
is reduced to computing these displacements.  If the elastic limits of
the contact are exceeded, the contact breaks or moves, and energy is
dissipated in this process.  In the following we describe the procedure
which is used in \sand to keep track of contact locations and
displacements.

When a new contact between grains $i$ and $j$ is formed, we define the
following vectors describing the position of the contact on each
grain.  Let $\vec{c}$ denote the position where the grain surfaces
touch.  In the program, this defines a new contact structure which is
kept in a list.  We define a vector $\RCPone=\vec{c}-\vec{r}_i$ which
points from the center of grain $i$ to the contact point and a vector
$\RCPtwo=\vec{c}-\vec{r}_j$ which points from the center of grain $j$
to the contact point.  We also define a unit vector \vec{Q} which is
perpendicular to the line connecting the two grain centers:
$|\vec{Q}|=1$ and $\vec{Q}\cdot (\vec{r}_i-\vec{r}_j) = 0$.
\begin{figure}[p]
\subfigure{\includegraphics[height=8cm]{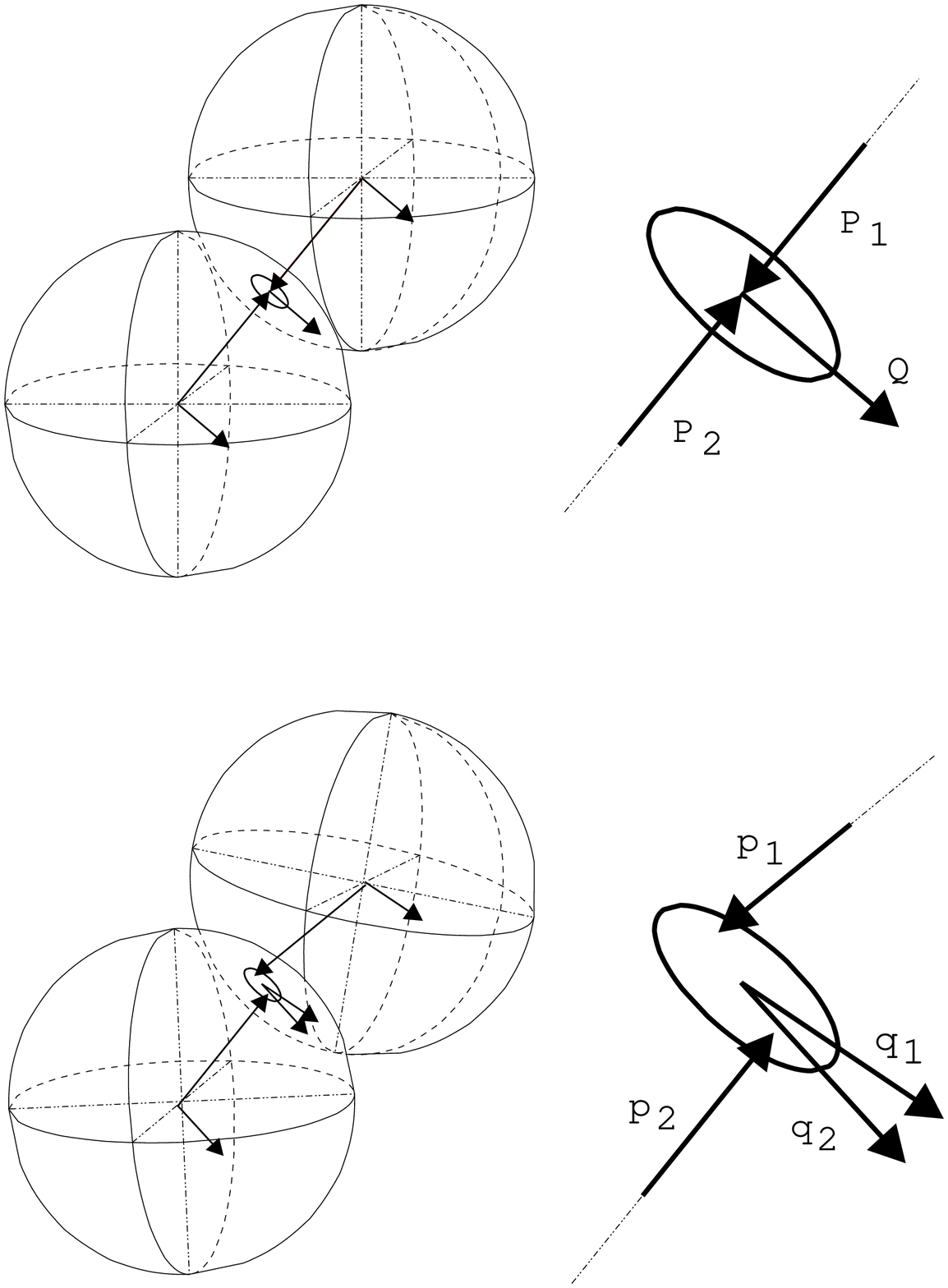}}\hspace{3cm}
\subfigure{\includegraphics[height=8cm]{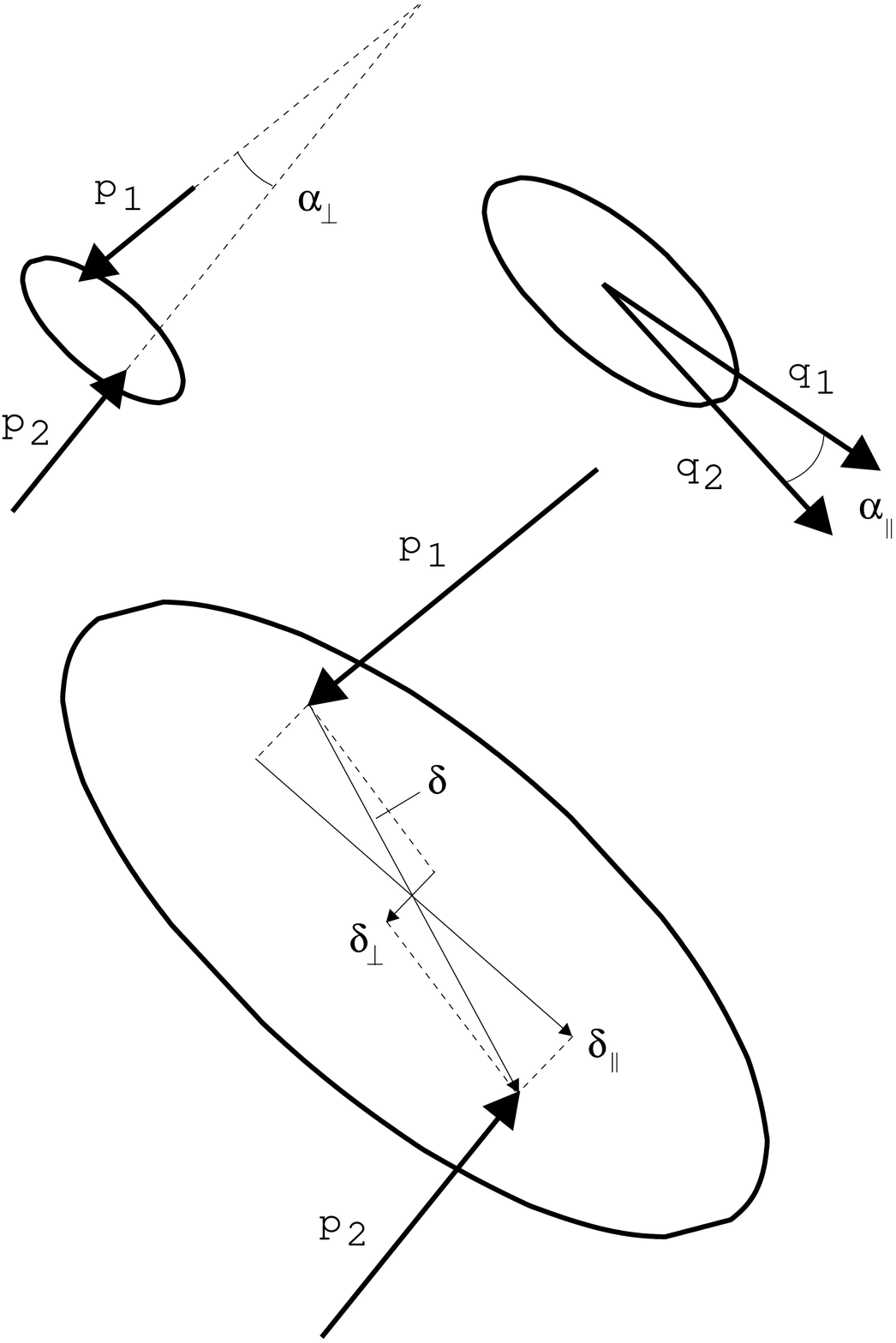}}
\caption{\label{fig:geometry}Geometrical treatment of contact
  displacements.  Left panel: Contact between two grains in
  equilibrium (top) and in stressed configuration (bottom).  Right
  panel: Blow-up of the contact region with definitions of the
  displacement vectors.  See text.}
\end{figure}
At a later stage during the calculation, the two grains $i$ and $j$
have moved so that their positions and/or rotational states are
different.  

If a contact were treated as being entirely rigid, the grains could
only move such that the contact remains in its original position on
the connecting line.  This corresponds to a treatment of aggregates
as rigid solid bodies \citep{richardson95}.  However, both grains are
still treated as individual particles and they can move in a different
way.  If the contact is not lost, the material near the contact is
strained in order to accommodate the small differences from rigid body
motion.  We can now use the vectors \RCPone, \RCPtwo, and \vec{Q} to
compute how much the material has been strained.
Figure~\ref{fig:geometry} shows the definition of these vectors.  The
displacements can then be calculated in the following way.  First we
rotate the vectors \RCPone, \RCPtwo, and \vec{Q} into the new
orientations of the grains, using the current rotation matrices
\begin{equation}
\vec{p}_1 = \rotmat_i \RCPone \quad,\quad
\vec{p}_2 = \rotmat_j \RCPtwo \quad,\quad
\vec{q}_1 = \rotmat_i \vec{Q} \quad,\quad
\vec{q}_2 = \rotmat_j \vec{Q} \mpkt
\end{equation}
The total linear displacement between the contact pointers is the vector
\begin{equation}
\label{eq:7}
\vec{\delta} = \vec{r}_j + \vec{p}_2 - \vec{r}_i - \vec{p}_1 \mpkt
\end{equation}
We define a unit vector pointing from the center of grain $j$ to
the center of grain $i$
\begin{equation}
\label{eq:6}
\vec{e}_{\perp} = \frac{\vec{r_1}-\vec{r_2}}{|\vec{r_1}-\vec{r_2}|}
\end{equation}
and compute the linear displacement in the direction of the inter-grain
axis $\vec{\delta}_{\perp}$ (pulling degree of freedom) and the
displacement in the contact plane $\vec{\delta}_{\parallel}$ (sliding
degree of freedom) with
\begin{align}
\vec{\delta}_{\perp} &= (\vec{\delta}\cdot\vec{e}_{\perp}) \vec{e_{\perp}} \\
\vec{\delta}_{\parallel} &= \vec{\delta} - \vec{\delta}_{\perp} \mpkt
\end{align}
Due to rolling of the surfaces over each other, the center of the
contact area is displaced from the inter-grain axis by
\begin{align}
\vec{\xi} &= \frac{(\vec{p}_1 - \vec{p}_2) \times
  (\vec{p}_1\times\vec{p}_2)}{(\vec{p}_1 - \vec{p}_2)^2} \mpkt
\end{align}
Finally we can compute the relative angular displacement around the
inter-grain axis $\alpha_{\parallel}$ (twisting degree of freedom) and
around an axis in the contact plane but perpendicular to the
$\vec{\xi}$ vector: $\alpha_{\perp}$ (rolling degree of freedom).
\begin{align}
\alpha_{\parallel} &= \arcsin(\vec{q}_{1}\times\vec{q}_2)\cdot \vec{e}_{\perp})\\
\alpha_{\perp} &= \arccos(\vec{p}_{1}\cdot(-\vec{p}_2)) \mpkt
\end{align}
 
From these displacements, the forces and torques acting on the grains
can be directly computed using the formalism described by
\citet{coagu}, with the following modifications:

\begin{itemize}
\item \cite{he99} and \cite{blum00:_exper_stick_restr} have shown that
  the critical rolling displacement $\xi_{\rm crit}$ is of the order
  of 10-20\AA{} instead of 1\AA{} as used by \citet{coagu}.  For the
  calculations in this paper, we have used $\xi_{\rm crit}=10$\AA.
  Rolling is the dominant restructuring process in the simulations in
  this paper, and a proper treatment is important here.  The
  measurements by Blum and Wurm have been restricted to Quartz
  particles, so the application to metallic or ceramic particles
  introduces some uncertainty.
  
\item \cite{poppe00:_aenal_exper_stick} have shown experimentally that
  the critical velocity for sticking can be considerably higher than
  the value predicted by \cite{Chokshi-ea93}.  This has to be
  interpreted as an additional channel of energy dissipation which is
  not yet well understood.  Since there is no good theoretical
  description available, we have still used the lower energies and
  velocities for sticking.  In the current study this effect is
  probably not of critical importance since  the attraction between
  grains is dominated by magnetic forces and generally leads
  to immediate sticking.
\end{itemize}

\subsection{Restructuring: making, breaking, moving contacts}

The algorithm described in the previous section is the one used while
contacts are in the elastic regime.  In this regime, the aggregate
keeps its basic shape except for minor bending and streching.  When the
dynamic forces acting on grains in contact become too big, the contact
will start to move over the two grains by rolling or sliding, or it
will break.  Within each time step, \sand assumes all contacts to be
stable and elastic, in order to make it possible for the integrator to
work on a set of continuous functions.  This assumption limits the
time step possible for \sand to the period of the fastest contact
vibration.  After each timestep, \sand checks if any of the contacts
are strained too much.  If this is the case and the grains are too far
apart to stay in contact, the contact is broken and removed from
the internal list.  If the grains are still close enough to touch, the
contact is moved to a new equilibrium location, releasing the stresses
that prompted the motion.  Then new values for the contact vectors
\RCPone, \RCPtwo, and \vec{Q} are computed and the time integration
continues.  Also after each time step, the integration volume is
checked for grains which have approached each other so that a new
contact has to be establised.

\subsection{List algorithm}

An $N$-particle code with long-range forces can only be numerically
efficient, if we can cut down the ${\cal O}\left( N^2
\right)$-dependence to a more acceptable ${\cal O}(N)$-dependence. To
within a numerical factor, this obviously represents the lower bound
for the amount of work required to process all $N$ particles.

In Eq.~\ref{e_rgg}, we defined a thermally induced interaction
limit for dipolar magnetic coupling. We can thus effectively reduce
the long-range interaction of a given dipole $j$ to that with a
restricted set of {\em neighbor} particles. The computational problem
is now reduced to implementing an efficient algorithm for list
construction and maintenance.  We have used a cell
subdivision \citep{ra95} adapted for the use with magnetically
interacting grains.

We use Eq.~\ref{e_rgg} as a definition for the {\em cut-off
  distance}, $r_c$, for our long-range forces. The simulation space is
subdivided into cells whose (identical) edges all exceed $r_c$ in
length. Interaction is only possible if two grains
are either in the same cell or in immediately adjacent (nearest neigh\-bors)
cells. In three dimensions, each cell has 26
neighboring cells. Due to reasons of symmetry only half of these need
be considered.
Note that this method only works properly if the size simulation
volume exceeds $4r_c$. Since we do not know in advance how many particles will occupy a given cell at
any instant, linked lists \citep{ra95} are used to associate particles
with the cell they reside in. 

The cell method can be
considerably refined by a neigh\-bor-list method. For cells with an edge
length $r_c$\,, on the average only a small fraction of the examined particles lie within interaction range. In order by
make use of this reduced neighborhood size, we have to construct a
list of interacting pairs from the set of particles found by the cell
method. To this end, we replace the cut-off distance $r_c$ by
$r_c+\Delta r$ with $\Delta r>0$ in order to allow the neighbor-list
to be useful during several numerical time steps, $\Delta t_i$. 
The concise value of $\Delta r$ is related to the rate at which the
list has to be updated in an inverse fashion. The list is updated whenever the
cumulative maximum velocity $\bv_{\rm max}$ exceeds a certain limit:
\begin{equation}\label{e_crit}
  \sum\limits_{\Delta t_i}\left|\bv_{i,\rm max}\right|>\frac{\Delta r}{\Delta t_i}\mpkt
\end{equation}
This is a rather conservative criterion since it sums contributions
from different particles. In practice, $\Delta r$-values between
0.3$r_c$ and 0.4$r_c$ have proved useful. For the evaluation according
to Eq.~\ref{e_crit} only unbound grains are considered. Bound
particles are subject to strong elastic forces \citep{coagu} that
cause rapid oscillation.

In order to include long-range forces on particles close to the edge
of the simulation volume, ``wraparound'' boundary conditions have been
implemented into \sand. 

For an ensemble of 100 magnetic particles, the combined cell-based
neigh\-bor-list method leads to a decrease in computation time by a
factor of 30 to 40. Since our approach is validated not only on a
numerical basis but also with respect to physical considerations we
have implemented an efficient way of dealing with magnetic forces in
an $N$-particle simulation.

\subsection{Multiple time scales}

Based on the considerations in the last section, dipolar
magnetic interaction of thermalized grains, i.e. grains that are
subject to random thermal motion and rotation, can be viewed as
moderately long-ranged with cut-off distance $r_c$. On these grounds
the {\em complete set} of $N(N-1)/2$ pair interactions can be divided
into a so-called {\em cardinal set} which contains all pairs $(i,j)$
for which $d_{ij}\leq r_c$, and the so-called {\em inessential set}
for which interactions are neglected \citep{st77}. The population of
both the cardinal and the inessential sets change with time.

The multiple time step (MTS) approach focuses on the
point that the cardinal set can be
subdivided into classes of {\em primary} and {\em secondary} neighbors
based on a distance criterion for potentials with strong $r$-dependence. It is
the rapidly changing force due to the primary neighbors that sets an
upper limit for the time-step $\Delta t$ in the simulation. In the MTS
method, only the contributions of these primary neighbors are updated
at every step. Longer time-steps are used to calculate the evolution of the secondary force, i.e.  the
contribution due to the secondary neighbor particles. To achieve this,
the secondary force
at time $t=t_0+k\Delta t$ is calculated from a Taylor series expansion based on
information at a given time $t_0$:
\begin{equation}\label{e_taylor1}
  \bF_s(t_0+k\Delta t) = \bF_s(t_0) + \sum\limits_{m}\bF_s^{(m)}(t_0)\;\frac{\left( k\Delta t  \right)^{(m)}}{m!}\mkomma
\end{equation}
where the superscript $(m)$ denotes the differentiation with respect
to time. A similar equation could be written down for the torque,
$\bM$. Equation~\ref{e_taylor1} is referred to as the $m$th-order Taylor
series. The maximum value $k_{\rm max}$ for the parameter $k$ has to
be chosen in such a way that the gain in computational efficiency
balances the loss in numerical accuracy. During the simulation, $k$
runs from 1 to $k_{\rm max}$. At that point a new Taylor series is
calculated according to Eq.~\ref{e_taylor1} and that step becomes
the new $t_0$.

While using additional memory for the storage of Taylor expansions,
this method can achieve an additional increase in computational speed
for all but the smallest systems up to a factor of five \citep{st77}.
The authors also report on the numerical error as a function of the
expansion order $m$ and the parameter $k_{\rm max}$.

In our coagulation model, an MTS approach is already inherent in the
numerical setup. In the scope of \sand, we identify the
elasto-mechanical contact forces with the primary force and the
magnetic forces with the secondary force. Since we use an adaptive
step size algorithm to integrate the equations of motion, the step
length $\Delta t$ is automatically adapted to the ``physical''
situation based on a prescribed absolute error criterion. As soon as
the first inter-particle contact is formed, the average step
size drops by a factor of 10$^2$ due to the strong elastic forces
acting on only a few particles. At this point we call upon the
MTS method, which has been implemented as an optional feature into
\sand.

Although straightforward mathematically, the calculation of the Taylor
expansion for magnetic dipole-dipole interaction is a rather
cumbersome task for
$\bF_{ij}=\bF_{ij}(\bmu_i,\bmu_j,\br_{ij} )$. As before,
the vector $\br_{ij}$ denotes the relative position of the two
dipoles involved. In what follows we shall drop the superscript $(ij)$
for simplicity's sake. Unless otherwise noted all vector quantities
are assumed to originate from particle $i$ and to be directed toward
or acting upon particle $j$. The basic formula involved is:
\begin{equation}\label{e_jj}
  \dot{\bF} = \underbrace{\frac{\partial \bF}{\partial \br}}_{:=\mat{J}_r\;\;}\cdot\frac{\partial \br}{\partial t}\;+\;\sum\limits_{k=i,j}\;\underbrace{\frac{\partial \bF}{\partial \bmu_k}}_{:=\mat{J}_{k,\mu}}\cdot\frac{\partial \bmu_k}{\partial t}\mkomma
\end{equation}
which is the total time derivative of the dipolar force. Since \sand
solves the full set of equations for particle motion including
rotation, the vectors $\dot{\bmu}_k$ are simply given by $ |
\bmu_k | \cdot \bomega_k$ if $\bomega_k$ denotes the
angular velocity of the grain. $\bv$ is the translational velocity of
the particle. The matrices $\mat{J}_r$ and $\mat{J}_{k,\mu}$
are the {\em Jacobian matrices}.

The Taylor expansion for the magnetic torque has to incorporate both
internal and external contributions. As before, we drop the
subscript $ij$. The expression for the first time derivative of
the torque $\dot{\bM}_j$ due to internal interaction then reads:
\begin{equation}\label{e_mdot}
  \dot{\bM}_{j,{\rm int}} = \frac{\partial \bmu_i}{\partial t}\times \bB + \bmu_i \times \frac{\partial \bB}{\partial t}\mpkt
\end{equation}
The expression for the dipolar magnetic field $\bB$ has been stated
before (cf. Eq.~\ref{e_mint2}). In analogy with Eq.~\ref{e_jj}, the time variation of the magnetic field, $\partial\bB / \partial
t$, can be determined from
\begin{equation}
  \dot{\bB} = \mat{J}_{r}\;\dot{\br} + \mat{J}_{i,\mu}\;\dot{\bmu}_i\mpkt
\end{equation}
The external contribution for each magnetic
grain is simply given by:
\begin{equation}\label{e_mdotext}
   \dot{\bM}_{j,{\rm ext}} =  |  \bmu_j  | \cdot \bomega_j \times \bB_{\rm ext}\mpkt
\end{equation}
We therefore use the following expression for the total time
derivative of the magnetic torque on a given particle $j$ which can be
derived by combining Eqs.~\ref{e_mdot}-\ref{e_mdotext}:
\begin{equation}\label{e_mdotcomp}
  \dot{\bM}_j = \dot{\bmu}_j\times\bB + \bmu_j\times\left( \mat{J}_r\;\dot{\br} + \mat{J}_{i,\mu}\;\dot{\bmu}_i\right) + |  \bmu_j  | \cdot \bomega_j \times \bB_{\rm ext}\mpkt
\end{equation}
In our numerical work, Taylor series expansions have been constrained
to first order due to the rather complicated structure of the
derivatives. These are recalculated at $t_0$ according to
Eqs.~\ref{e_jj} and \ref{e_mdotcomp}. During the following $k_{\rm
  max}$ time steps, which are governed by the high numerical demands
of strong contact forces, all magnetic forces are calculated using
Eq.~\ref{e_taylor1} and its counterpart for the torques. In
practice, the choice $k_{\rm max}=70$ has yielded satisfactory results
both in terms of computing time and physical accuracy.

\bibliographystyle{natbib-icarus}
\bibliography{ms}

\end{document}